\documentclass[12pt, prd, showpacs]{revtex4}
\usepackage{amssymb}

\input{tcilatex}

\begin{document}

\title{Boulware state in exactly solvable models of 2D dilaton gravity}
\author{O. B. Zaslavskii}
\affiliation{Department of Mechanics and Mathematics, Kharkov V.N.Karazin National
University, \\
Svoboda Square 4, Kharkov 61077, Ukraine}
\email{ozaslav@kharkov.ua}

\begin{abstract}
We discuss self-consistent geometries and behavior of dilaton in exactly
solvable models of 2D dilaton gravity, with quantum fields in the Boulware
state. If the coupling $H(\phi )$ between curvature and dilaton $\phi $ is
non-monotonic, backreaction can remove the classical singularity. As a
result, an everywhere regular star-like configuration may appear, in which
case the Boulware state, contrary to expectations, smooths out the system.
For monotonic $H(\phi )$ exact solutions confirm the features found before
with the help of numerical methods: the appearance of the bouncing point and
the presence of isotropic singularity at the classically forbidden branch of
the dilaton.

\end{abstract}

\maketitle



\section{Introduction}

The backreaction of quantum field on a black hole is significant at late
stages of evaporation due to the Hawking effect and, thus, affects a final
fate of a black hole. As far as a static black hole is concerned, the role
of backreaction on the background of a non-extremal black hole depends
crucially on the state of quantum fields. In the Hartle-Hawking state the
quantum stresses are bounded everywhere, so that small backreaction shifts
parameters of a black hole but does not change its properties qualitatively.
The situation changes drastically for the Boulware state that corresponds to
the Minkowski vacuum at infinity. As is well known, near the horizon such
quantum stresses diverge, so in general relativity a classical regular
horizon is destroyed in this case. The question arises, what happens to the
quantum-corrected geometry whose classical limit corresponds to a black
hole. In doing so, one cannot consider the behavior of quantum stresses in a
fixed background but, instead, should solve field equations in a
self-consistent way.

Motivation to study such geometries comes from different directions. As is
pointed out in \cite{fab1}, \cite{fab2} this concerns analogy with
brane-world physics in the 5D world, where the problem of finding static
black hole solutions is mapped to the problem of finding self-consistent
solutions in the 4D world with quantum fields in the Boulware state \cite%
{5d1}, \cite{5d2}. Another line of motivation is, in our view, connected
with the physically important problem of the fate of Cauchy horizons which
are present, say, in classical charged black holes. It is believed that it
is classically unstable (see, e.g. Sec. 14.3.2 of the monograph \cite{fn}
and references therein). Meanwhile, one can expect also another kind of
instability here due to infinite quantum backreaction of massless fields
near such horizons. This resembles the situation with the Boulware state but
the latter is certainly more simple, so its study could help us to
understand better what happens in the vicinity of Cauchy horizons. Apart
form this, the Boulware state as such arises in \ rather simple and natural
way, being singled out by the condition that the quantum stresses vanish at
infinity, so one is led to elucidate what happens to self-consistent
solutions whose classical analogues were black holes. In particular, it is
not clear in advance whether we will obtain a singular horizon or regular
configuration without a horizon. This probelms

Recently, the problem under discussion for the Schwarzschild black hole was
tackled in \cite{fab1}, \cite{fab2} where it was found that the metric
component $g_{00}$ and the areal radius cease to be monotonic functions of
the coordinate and acquire bouncing points. Similar features seem to occur
also in higher-dimensional theories. These conclusions alter the general
picture qualitatively in that quantum-corrected geometries whose classical
counterparts were singular near a classical would-be horizon, become regular%
\textit{\ }there but generate a branching point beyond it and a singularity
having no classical analogue.

Meanwhile, the validity of the results obtained in \cite{fab1}, \cite{fab2}
is obscured by the assumptions that cannot be controlled easily. The
corresponding approach exploits spherically-symmetrical reduction from the
four-dimensional (4D) theory to the effective two-dimensional (2D) one. In
doing so, all contributions to the stress-energy tensor except the s-wave
one are ignored that, in general, is not quite good approximation \cite{fz}.
Further, simplification was used in \cite{fab1}, \cite{fab2} based on the
replacement of the action of scalar field by its Polyakov-Liouville form. It
was motivated by the properties of the wave equation near the usual
non-extremal regular horizon. However, it is not quite clear how this
approximation and the distinction between 2D and 4D theories as a whole
affect the results in the entire region and, in addition, the horizon itself
can become singular. Even after all these approximations the field equations
remain quite complicated and one is led to the implementation of numerical
methods.

Fortunately, there is the situation when field equations admit
self-consistent exact solutions with backreaction taken into account. This
is 2D dilaton gravity considered as a true 2D theory from the very
beginning, with minimal quantum fields described by the Polyakov-Liouville
action without any additional assumptions. Remarkably, such theories admit
exactly solvable models provided the action coefficients obey some relation.
In the present paper we show some qualitative features of the Boulware state
using simple and known exactly solvable models and their generalizations. In
particular, we trace how backreraction can destroy classical horizons and/or
singularities and create new ones of pure quantum nature. Some results turn
out to be similar to those found in \cite{fab1}, \cite{fab2} while some seem
not to have counterparts in general relativity (but may have them in 4D
dilaton gravity). In a more formal way, one can pose the problem of finding
self-consistent solutions of field equations under two conditions: (i)
quantum stresses vanish at infinity, (ii) in the classical limit there
exists a classical horizon. Then we will see that there are two main cases:
either the singular horizon exists (in which case we consider the type of
singularity in more detail) or the horizon disappears at all giving way to a
star-like configuration. (We must make reservation that we do not consider
black hole solutions in higher curvatures theories - see, e.g., \cite{4}.)

\section{Basic equations}

Consider the action 
\begin{equation}
I=I_{0}+I_{PL}\text{,}  \label{action}
\end{equation}%
where 
\begin{equation}
I_{0}=\frac{1}{2\pi }\int_{M}d^{2}x\sqrt{-g}[F(\phi )R+V(\phi )(\nabla \phi
)^{2}+U(\phi )]  \label{clac}
\end{equation}%
and $I_{PL}$ is the Polyakov-Liouville action incorporating effects of
quantum fields minimally coupled to the background. We assume that
backreaction is due to a multiplet of N fields and neglect quantum
backreaction of dilaton itself in the large N approximation, as usual (see.
e.g. Sec. 3.7 of \cite{str}). In doing so, 
h{\hskip-.2em}\llap{\protect\rule[1.1ex]{.325em}{.1ex}}{\hskip.2em}%
$\rightarrow 0\,\ N\rightarrow \infty $ in such a way that the quantum
coupling parameter $\kappa =\frac{\text{%
h{\hskip-.2em}\llap{\protect\rule[1.1ex]{.325em}{.1ex}}{\hskip.2em}%
}N}{24}$ remains finite. Let us consider the potential in the form $%
U=4\lambda ^{2}\exp (-2\phi )$ that arises naturally in string-inspired
models. If the action coefficients obey the condition%
\begin{equation}
V=-2(\frac{dF}{d\phi }+\kappa )\text{,}
\end{equation}%
the field equation have exact static solutions which admit complete
classification \cite{class} (see also references therein). Then we may
borrow directly explicit expressions from eqs. (52), (53) of \cite{class}
where the case of Boulware state for the asymptotically flat metric is
contained implicitly among a variety of exact solutions. Hereafter, we use
the quantum-corrected coefficient $H(\phi )=F-\kappa \ln U$ instead of the
original one $F(\phi )$. In conformal coordinates the metric takes the form%
\begin{equation}
ds^{2}=g(-dt^{2}+d\sigma ^{2})\text{,}  \label{conf}
\end{equation}%
\begin{equation}
R=-\frac{2\lambda ^{2}}{g}\frac{d^{2}\phi }{dy^{2}}\text{, }y\equiv \lambda
\sigma  \label{rf}
\end{equation}%
and we have 
\begin{equation}
H(\phi )=f(y)\text{, }f(y)\equiv e^{2y}-\kappa y+H_{0}\text{, }g=e^{2y+2\phi
}\text{.}  \label{f}
\end{equation}%
\begin{equation}
T_{1}^{1(PL)}=-\frac{1}{4\pi }e^{-2y}\{2\alpha +\kappa (\frac{\alpha }{%
2\lambda }-2\lambda e^{2y})2H^{^{\prime }-1}[-4\lambda +(\frac{\alpha }{%
2\lambda }-2\lambda e^{2y})2H^{^{\prime }-1}]\}\text{,}  \label{11}
\end{equation}%
Here $R$ is the Riemann curvature, $T_{\nu }^{\mu (PL)}$ is the
stress-energy of quantum fields, $\kappa =\frac{N}{24}$ is the quantum
coupling parameter, $N$ is a number of field in a multiplet. As is well
known, the trace $T_{\mu }^{\mu (PL)}=\kappa \frac{R}{\pi }$. We also assume
that at the right infinity $y\rightarrow \infty $ there is a linear dilaton
vacuum, $\phi =-y$, $g\rightarrow 1$. Correspondingly, we assume that,
asymptotically at $\phi \rightarrow -\infty $, $H\simeq \exp (-2\phi )$. The
Boulware state implies that $T_{1}^{1(PL)}\rightarrow 0$ at $y\rightarrow
\infty $. It is easy to check by substitution into (\ref{11}) that this is
achieved by the choice $\alpha =2\lambda ^{2}\kappa $.

In the classical limit $\kappa =0$. Then, at $y\rightarrow -\infty $ and
finite $\phi =\phi _{h}$ we have $g\rightarrow 0$, so this is a black hole
horizon, $H_{0}=H(\phi _{h})$. If $H=e^{-2\phi }$, we obtain an usual CGHS
black hole \cite{cg} with $g=\frac{e^{2y}}{e^{2y}+H_{0}}=1-e^{2(x_{h}-x)}$,
hereafter $x=\int dyg$ is the Schwarzschild-like coordinate, $x_{h}$ is its
horizon value.

It is essential that, whatever the function $H(\phi )$ be, the function $%
f(y) $ has an universal form, independent of the particular dilaton gravity
model. At $y\rightarrow \infty $ and $y\rightarrow -\infty $ the function $%
f\rightarrow \infty $. It has one local minimum at $y=y_{0}=\frac{1}{2}\ln 
\frac{\kappa }{2}$,$\ f(y_{0})=H_{0}+\frac{\kappa }{2}(1-\ln \frac{\kappa }{2%
})$.

\section{Behavior of metric and dilaton}

In what follows we will be interested only in such configurations when the
classical limit ($\kappa =0$) corresponds to a black hole. This excludes the
situation when $H(\phi _{0})>f(y_{0})$ where $\phi _{0}$ is the point of
minimum of $H$ (if it exists). Indeed, in this case, when $y$ attains the
point $y_{1}$ where $f(y_{1})=H(\phi _{0})$, eq. (\ref{f}) gives us that $%
H^{\prime \prime }(\phi _{0})\frac{(\phi _{0}-\phi )^{2}}{2}=f^{\prime
}(y_{1})(y-y_{1})$, the metric function $g$ is finite, the curvature $R\sim
H^{\prime -3}\sim (y-y_{1})^{-\frac{3}{2}}$ diverges, so we have a naked
time-like singularity which persists also in the limit $\kappa \rightarrow 0$%
. The horizon does not exist at all, so this situation has nothing to do
with the Boulware state. In a similar way, if $H$ approaches its limiting
value at $\phi \rightarrow \infty $ asymptotically and $H(\infty )=f(y_{0})$%
, a semi-infinite throat arises instead of \ a black hole \cite{bose} that
is beyond the scope of our paper. Therefore, we restrict ourselves by the
following cases.

1) $H(\phi )$ is monotonic and unbounded. The typical example is

\begin{equation}
H=e^{-2\phi }-a\kappa \phi \text{, }U=4\lambda ^{2}e^{-2\phi }\text{, }a>0%
\text{.}  \label{m1}
\end{equation}

The dilaton field $\phi =-\infty $ at $y=\infty $. If, further, $y$
decreases, classically the dilaton field grows monotonically up to the
horizon where $y=-\infty $ and $\phi =\phi _{h}$ is finite, $H(\phi
_{h})=H_{0}$. However, if $\kappa \neq 0$, the bouncing point appears in the
solution at $y=y_{0}$. At this point the derivative $\frac{d\phi }{dy}$
changes its sign, so $\phi $ again decreases to $-\infty $ as $y\rightarrow
\infty $. At the point $y_{0}$ the metric function $g(y_{0})=\frac{\kappa }{%
U(y_{0})}$. The fact that $\left( \frac{d\phi }{dy}\right) _{y=y_{0}}=0$
while the metric function $g\neq 0$ entails that $\left( \nabla \phi \right)
^{2}=0$. This is nothing other than the 1+1 analogue of the apparent horizon 
$(\nabla r)^{2}=0$ which does not coincide now with the event horizon. In
the limit $\kappa \rightarrow 0$ we have $y_{0}\rightarrow -\infty $, $%
g(y_{0})\rightarrow 0$ and a classical event horizon restores. For a given
small but non-zero value of $\kappa $ the whole region $y<y_{0}$ has no
classical counterpart.

The metric function in (\ref{f}) can be rewritten also in the form $%
g=e^{2\rho }$, $\rho =y+\phi $. At $y=y_{0}$ $\frac{d\rho }{dy}=1$, and we
have $\left( \frac{dr}{d\rho }\right) _{y=y_{0}}=0$, where $r=\exp (-\phi )$
also in qualitative agreement with \cite{fab1}, \cite{fab2}.

At $y\rightarrow -\infty $ the dilaton field for the model (\ref{m1}) $\phi
=-\frac{1}{2}\ln \left\vert y\right\vert $, the metric function $g\sim \frac{%
\exp (2y)}{\left\vert y\right\vert }\rightarrow 0$, the curvature 
\begin{equation}
R\sim -\frac{\lambda ^{2}\kappa ^{2}}{r^{2}}\exp (2\frac{r^{2}}{\kappa })=-%
\frac{\lambda ^{2}\kappa }{\left\vert y\right\vert }\exp (2\left\vert
y\right\vert )\text{,}  \label{r1}
\end{equation}%
where $r\rightarrow \infty $, so we have a singular horizon. Thus, the
non-zero quantum coupling parameter $\kappa $ is essential in that $%
\lim_{y\rightarrow -\infty }R$ diverges. The features of the solution
discussed above in case 1) are qualitatively similar to those found in \cite%
{fab1}, \cite{fab2}.

2) $H(\phi )\rightarrow \infty $ at $\phi \rightarrow \pm \infty $ and has
one minimum at $\phi =\phi _{0}$. The typical example is 
\begin{equation}
H=e^{-2\phi }+c\phi \text{, }U=4\lambda ^{2}e^{-2\phi }\text{, }c=b\kappa >0%
\text{.}  \label{b}
\end{equation}%
The value $b=1$ corresponds to the so-called RST\ model \cite{rst1}, \cite%
{rst2}. For a given $y$, there exist two branches of the solution $\phi (y)$%
. For definiteness, we choose the branch that corresponds to the linear
dilaton vacuum $\phi =-y$ at the right infinity. Then, depending on the
value of $H_{0}$, we have the following subcases.

a) $H(\phi _{0})<f(y_{0})$. Then, as $y$ changes from plus to minus
infinity, the point representing our system on the curve $H(\phi )$ moves
from $\phi =-\infty $ at $y=\infty $ until the point $y_{0}$ and back to $%
\phi =-\infty $ for $y<y_{0}$ along the same branch of the function $H(\phi
) $, so the situation is similar to case 1, the presence of the second
branch of $H(\phi )$ does not manifest itself.

b) The most interesting case arises when $\phi _{0}=\phi (y_{0})$, so that $%
H(\phi _{0})=f(y_{0})$. Then near $y_{0}$ 
\begin{equation}
H^{\prime \prime }(\phi _{0})\frac{(\phi -\phi _{0})^{2}}{2}=f^{\prime
\prime }(y_{0})\frac{(y-y_{0})^{2}}{2}\text{,}  \label{hf}
\end{equation}%
the derivative $\frac{d\phi }{dy}=\frac{f^{\prime }(y)}{H^{\prime }}=\frac{%
f^{\prime \prime }(y_{0})(y-y_{0})}{H^{\prime \prime }(\phi -\phi _{0})}%
\rightarrow -\sqrt{\frac{f^{\prime \prime }(y_{0})}{H^{\prime \prime }(y_{0})%
}}<0.$Thus, the turning point for the dilaton disappears, near $y_{0}$ the
dilaton can be expanded into series with respect to $y-y_{0}$, the metric
function $g$ and curvature being finite.

It is worth reminding that, for a given $H_{0}$, the singularity of the
classical CGHS black hole was situated at $\phi \rightarrow \infty $, $%
x\rightarrow -\infty $ $\ $(the proper distance being finite), $R\sim \exp
(-x)$. Inclusion of backreaction shifts the singularity to the finite value $%
\phi _{0}$. It was shown in \cite{solod} that in the Hartle-Hawking state of
the RST model the curvature $R\sim (x_{h}-x)^{-3/2}$, if (in our notations) $%
H(\phi _{0})<f(y_{0})$. The singularity becomes milder, $R\sim
(x_{h}-x)^{-1/2}$, if $H(\phi _{0})=f(y_{0})$, but still persists.
Meanwhile, as we see, in the Boulware state the singularity at $\phi =\phi
_{0}$ disappears at all. In this sense, contrary to expectations, the
Boulware turns out to be "more regular" than the Hartle-Hawking one!

Thus, $\phi (y)$ continues to diminish when we pass the point $\phi _{0}$ in
the direction of increasing the dilaton field and move to the branch $\phi
>\phi _{0}$.Then, we have from (\ref{f}) and (\ref{b}) that, asymptotically
at $y\rightarrow -\infty $ the dilaton $\phi \sim -\frac{y}{b}$ and 
\begin{equation}
g\sim \exp [2\left\vert y\right\vert (\frac{1}{b}-1)]\text{, }R\sim \exp
[-2\left\vert y\right\vert q]\text{, }  \label{gb}
\end{equation}%
where $q=\frac{1}{b}$ for $b\leq 1$ and $q=\frac{2}{b}-1$ for $b\geq 1$. The
properties of the spacetime depend crucially on the value of $b$.

b1) $b<1$. Then, $g\rightarrow \infty $, the curvature $R\rightarrow 0$. In
terms of the Schwarzschild-like coordinate the metric function $g\sim
\left\vert x\right\vert $. Thus, the metric has the asymptotically Rindler
form and, in this sense, corresponds to an accelerated observer. However,
there is no sense to speak about the acceleration horizon at $x=0$ since
this form of the metric is valid for $\left\vert x\right\vert \rightarrow
\infty $ only. One can check that in (\ref{11}) $T_{1}^{1(PL)}\rightarrow 0$
because of the factor $g^{-1}\rightarrow 0$.

b2) $b=1$. This is the RST model \cite{rst1}, \cite{rst2}. Then $%
g\rightarrow const$, $R\rightarrow 0$, so we have the Minkowski spacetime at
both infinities. Now, $T_{1}^{1(PL)}\rightarrow 0$ because of vanishing
braces in (\ref{11}).

b3) $1<b\leq 2$. Then $g\rightarrow 0$ (horizon) and $R\rightarrow 0$ or $%
R\rightarrow const$. The remarkable feature of these solutions consists in
that the geometry near the horizon is regular but quantum stresses $%
T_{1}^{1(PL)}$ diverge. There exists the whole class of such models
described in more detail in \cite{found2}.

b4) $b>2$. Then $g\rightarrow 0$ but $R\rightarrow \infty $, so we have a
singular horizon.

Up to now, the parameter $\kappa $ played the double role in that it changed
the behavior of both functions $H(\phi )$ and $f(y)$. For the string
inspired models like (\ref{b}) in the classical limit $\kappa =0$ the
function $H=e^{-2\phi }$ is monotonic but $H$ has a minimum for $\kappa \neq
0$. In a similar way, the function $f(y)=e^{2y}+H_{0}$ is monotonic
classically but acquires the minimum at $y_{0}$ when $\kappa \neq 0$.
Meanwhile, it makes sense to consider a somewhat different situation in
which the parameter $c>0$ in (\ref{b}) has a non-vanishing classical limit
(for example, we may take $c=d+\kappa b$ where $d$ does not contain $\kappa $%
). Then, account for backreaction changes the behavior of $f(y)$
qualitatively but only slightly changes the value of the coefficient $c$ in $%
H(\phi )$. This allows us to compare the properties of the Boulware state to
those of the classical configuration in a more direct way. As case 2a) is
similar to case 1) with the monotonic $H(\phi )$, we restrict ourselves to
case 2b).

If the condition $H(\phi _{0})=f(y_{0})$ is satisfied, in the classical
limit $y_{0}\rightarrow -\infty $ and eq. (\ref{f}) gives rise to $H^{\prime
\prime }(\phi _{0})\frac{(\phi -\phi _{0})^{2}}{2}=e^{2y}\sim x-x_{h}$,
whence $\phi =\phi _{0}-\sqrt{\frac{2}{H^{\prime \prime }(\phi _{0})}}%
e^{y}[1+O(e^{y})]$. As a result, the curvature $R\sim e^{-y}\sim
(x_{h}-x)^{-1/2}$ like in the Hartle-Hawking state of the RST model \cite%
{solod}. In the quantum case the behavior of the dilaton is determined near $%
y_{0}$, in the main approximation, by a quite different equation (\ref{hf}).
Keeping also the third terms of the Taylor expansion near $\phi _{0}$ one
finds that $3\frac{d^{2}H}{d\phi ^{2}}(\phi _{0})\left( \frac{d^{2}\phi }{%
dy^{2}}\right) _{y=y_{0}}=\left( \frac{d\phi }{dy}\right) _{y=y_{0}}^{-1}%
\frac{d^{3}f(y_{0})}{dy^{3}}-\frac{d^{3}H}{d\phi ^{3}}(\phi _{0})\left( 
\frac{d\phi }{dy}\right) _{y=y_{0}}^{2}$. Taking into account that $\frac{%
d^{2}f(y_{0})}{dy^{2}}\sim \frac{d^{3}f(y_{0})}{dy^{3}}\sim \kappa $ while $%
\frac{d^{2}H}{d\phi ^{2}}(\phi _{0})$ and $\frac{d^{3}H}{d\phi ^{3}}(\phi
_{0})$ do not contain $\kappa $, $g\sim e^{2y_{0}}\sim \kappa $, we obtain
from (\ref{rf}) that the curvature is finite due to backreaction and in the
limit $\kappa \rightarrow 0$ $R(y_{0})\sim \frac{1}{\sqrt{\kappa }}$.

Near $y\rightarrow \infty $ eq. (\ref{gb}) is valid in which we must replace 
$b\rightarrow \frac{c}{\kappa }$. If $c$ is finite but $\kappa \rightarrow 0$%
, $b\gg 1$, so that case b4) is realized and a singular horizon appears.

To conclude this section, we would like to point out that the coupling of
dilaton to curvature is described in the quantum-corrected action by the
coefficient $H$ instead of $F$ in the pure classical case (\ref{clac}).
Therefore, the effective gravitation-dilaton constant $\gamma \sim H^{-1}$.
As a result, it follows from (\ref{m1}), (\ref{b}) that even in the region
where the dilaton $\phi $ itself infinitely grows, this constant remains
finite and, moreover, tends to zero like $\phi ^{-1}$. By contrast, in the
pure classical action the corresponding constant $\gamma \sim F^{-1}$
infinitely grows. In this sense, account for backreaction automatically
expands the region of validity of semiclassical approximation.

\section{Summary}

The advantage of the approach under discussion is that we solve the problem
analytically and, moreover, have at our disposal exact solutions which are
quite simple and convenient for analysis. We considered two main situations.
1) The curvature coupling $H(\phi )$ is monotonic. Then we demonstrated such
features as the appearance of the bouncing point in the solution,
transformation of the classical regular event horizon into the apparent one
and the appearance of the singular horizon having no classical analogue. All
these features are qualitatively similar to those found in \cite{fab1}, \cite%
{fab2}, although the concrete gravitational-dilaton actions do not coincide.

However, there is also a quite different case 2b) which seems to have no
analogue in general relativity and in the effective 2D theory considered in 
\cite{fab1}, \cite{fab2}. If $H(\phi )$ has a minimum $\phi _{0}$ where $%
H(\phi _{0})=f(y_{0})$, the singular horizon typical of the Hartle-Hawking
state or the classical black hole disappears in the Boulware state, the
point $\phi _{0}$ being regular due to backreaction. The new part of the
manifold opens behind it which, classically, was unreachable from infinity.
In doing so, there is neither an apparent nor an event horizon in the
vicinity of the point $y_{0}$. A singular horizon may appear to the left
from this point but it is situated at a finite distance from it and has pure
quantum origin. In doing so, the dilaton value $\phi \rightarrow \infty $
while the typical classical horizon has a finite value $\phi _{h}$ on the
horizon. Apart from this, the region beyond the point $y_{0}$ may represent
a star-like configuration with the asymptotic left infinity of two kinds:
either the Minkowski metric or the Rindler one. In addition, we would like
to pay attention to the previous observation \cite{found1}, \cite{found2}
that quantum stresses in the Boulware state that destroy the classical
would-be horizon completely may also lead to the appearance of a new regular
(!) horizon of pure quantum origination. We see that the Boulware state may
not only bring about singularities into a classical black hole background
but, vice versa, smooth out the singularities of the classical counterpart
and create everywhere regular configurations.

Thus, quantum terms not simply dominate near the classical would-be horizon
but may change the structure of the spacetime qualitatively. It is of
interest to trace whether similar effects can occur in 4D dilaton theory.

\end{document}